\documentclass[english,authoryear]{elsarticle}
\usepackage[T1]{fontenc}
\usepackage[latin9]{inputenc}
\usepackage{amsmath}
\usepackage{amssymb}
\usepackage{graphicx}
\usepackage{esint}


\usepackage{babel}

\usepackage{lineno}

\journal{BioSystems}

\usepackage{babel}

\usepackage{amsthm}

\usepackage{hyperref}

\makeatother

\usepackage{babel}
\begin{document}

\title{Quantum-like model of behavioral response computation using neural
oscillators}

\author{J. Acacio de Barros}

\address{Liberal Studies Program, San Francisco State University, San Francisco,
CA 94132}
\begin{abstract}
In this paper we propose the use of neural interference as the origin
of quantum-like effects in the brain. We do so by using a neural oscillator
model consistent with neurophysiological data. The model used was
shown elsewhere to reproduce well the predictions of behavioral stimulus-response
theory. The quantum-like effects are brought about by the spreading
activation of incompatible oscillators, leading to an interference-like
effect mediated by inhibitory and excitatory synapses. \end{abstract}
\begin{keyword}
disjunction effect, quantum cognition, quantum-like model, neural
oscillators, stimulus-response theory
\end{keyword}
\maketitle

\section{Introduction}

Quantum mechanics is one of the most successful scientific theories
in history. From it, detailed predictions and extremely precise descriptions
of a wide range of physical systems are given. Thus, it is not surprising
that researchers from social and behavioral sciences disciplines started
asking whether the apparatus of quantum mechanics could be useful
to them \citep{khrennikov_ubiquitous_2010}. In applying the quantum
formalism, most researchers do not claim to have found social or behavioral
phenomena that are determined by the physical laws of quantum mechanics
\citep{bruza_introduction_2009}. Instead, they claim that by using
the representation of states in a Hilbert space, together with the
corresponding non-Kolmogorovian interference of probabilities, a better
description of observed phenomena could be achieved. To distinguish
the idea that the formalism describing a system is quantum without
requiring the system itself to be, researchers often use the term
\emph{quantum-like. }

One of the first applications of quantum-like dynamics outside of
physics was in finances \citep{baaquie_path_1997,haven_discussion_2002}
and in decision making \citep{busemeyer_quantum_2006,aerts_quantum_2009}.
In economy, \citet{baaquie_path_1997} showed that a more general
form of the Black-Scholes option pricing equation was given by a Schrödinger-type
equation, and \citet{haven_discussion_2002} linked the equivalent
of the Plank constant to the existence of arbitrage in the model.
In psychology, most of the initial effort was focused on using quantum-like
dynamics and Hilbert-space formalisms to describe discrepancies between
empirical data and ``classical'' models using standard probability
theory (see, for example, \citet{aerts_quantum_2009,asano_quantum-like_2010,busemeyer_quantum_2006,busemeyer_quantum_2007,busemeyer_empirical_2009,khrennikov_quantum-like_2009,khrennikov_quantum_2009,khrennikov_quantum-like_2011}).

To try to describe quantum-like effects in cognitive sciences, in
a recent paper \citet{khrennikov_quantum-like_2011} used a classical
electromagnetic-field model for the mental processing of information.
In it, Khrennikov argues that the collective activity of neurons involved
in mental processing produces electromagnetic signals. Such signals
then propagate in the brain, leading to interference and the appearance
of quantum-like correlations. He used this model to show how quantum-like
effects help understand the binding problem. 

In this paper we set forth a neural oscillator model exhibiting many
of the characteristics of Khrennikov's. The neural model we employ
was initially proposed as an attempt to obtain behavioral stimulus-response
theory from the neuronal activities in the brain \citep{suppes_phase-oscillator_2012}.
There are many different ways to use neural oscillators to model learning
\citep[see ][ and references therein]{vassilieva_learning_2011},
but to our knowledge, this is the only one that makes a clear connection
between neural oscillator computations and the behavioral responses.
Furthermore, the model is simple enough to help understand the basic
principles behind the computation of a response from a stimulus without
compromising the neurophysiological interpretation of it. This must
be emphasized, as most neural models constructed from a neurophysiological
basis are so computationally intensive and complex that it is very
hard to understand why or how one response was select instead of another.
We shall see later that this is not the case with \citet{suppes_phase-oscillator_2012}. 

As with Khrennikov's electromagnetic field model, our model has many
quantum-like characteristics, such as signaling, superposition, and
interference. However, it does not rely on the propagation of electromagnetic
fields in the brain. Instead, we show that these quantum-like effects
could be obtained by directly considering the interference of the
activities of collections of firing neurons coupled via inhibitory
and excitatory synapses. In particular, we show a violation of Savage's
sure-thing principle, often cited as an example of quantum-like behavior. 

This paper is organized as follows. In section \ref{sec:SR-theory-neural-oscillators}
we introduce Suppes, de Barros, and Oas's \citeyearpar{suppes_phase-oscillator_2012}
oscillator model, with special emphasis to its neurophysiological
interpretation and connection to SR theory. Then, in section \ref{sec:Quantum-like-effects},
we present a contextual theory of quantum-like phenomena coming from
classical interference of waves \citep{de_barros_quantum_2009}. Finally,
in section \ref{sec:SR-theory-neural-oscillators} we show computer
simulations of neural oscillators and how they fit some quantum-like
experimental data in the literature. We end with some remarks.

\section{Neural oscillator model of SR-theory \label{sec:SR-theory-neural-oscillators}}

In this section, we delineate the modeling of SR theory by neural
oscillators. This section is divided into three parts. First, in subsection
\ref{sub:Review-of-SR} we sketch the mathematical version of behavioral
SR theory we model. Then, in subsection \ref{sub:Qualitative-description-of}
we present the model of \citet{suppes_phase-oscillator_2012} in a
qualitative way. Finally, in subsection \ref{sub:Quantitative-description-of}
we give a mathematical description of the oscillator model. Our goal
is not to be exhaustive, but only to focus on the main features of
the model that are relevant for the emergence of quantum-like effects
derived later. Readers interested in more detail are referred to \citep{suppes_phase-oscillator_2012,de_barros_response_2012}.

\subsection{Review of SR theory\label{sub:Review-of-SR}}

Stimulus-response theory (or SR theory) is one of the most successful
behavioral learning theories in psychology \citep{suppes_markov_1960}.
Among the many reasons for its success, it has been shown to fit well
empirical data in a variety of experiments. This fitting requires
few parameters (the learning probability $c$ and the number of stimuli),
giving it strong prediction powers from few assumptions. Also, because
it requires a rigid trial structure, its concepts can be formally
axiomatized, resulting in many important non-trivial experiments.
Finally, despite its simplicity, SR theory is rich enough such that
even language can be represented within its framework \citep{suppes_representation_2002}.
Thus, SR theory presents an ideal mathematical structure to be modeled
at a neuronal level%
\footnote{SR theory, in the form presented here, has lost most of its support
among learning theorists, mainly because of of inability to satisfactorly
explain certain experiments \citep{mackintosh_conditioning_1983}.
However, many of its theorems are relevant to current neurophysiological
experiments. %
}. 

Here we present the mathematical version of SR theory for a continuum
of responses, formalized in terms of a stochastic process \citep[here we follow ][]{suppes_phase-oscillator_2012}.
Let $\left(\Omega,{\cal F},P\right)$ be a probability space, and
let $\mathbf{Z}$, $\mathbf{S}$, $\mathbf{R}$, and $\mathbf{E}$
be random variables, with $\mathbf{Z}:\Omega\rightarrow E^{\left|S\right|}$
$\mathbf{S}:\Omega\rightarrow S$, $\mathbf{R}:\Omega\rightarrow R$,
and $\mathbf{E}:\Omega\rightarrow E$, where $S$ is the set of stimuli,
$R$ the set of responses, and $E$ the set of reinforcements. Then
a trial in SR theory has the following structure: 
\begin{equation}
\mathbf{Z}_{n}\rightarrow\mathbf{S}_{n}\rightarrow\mathbf{R}_{n}\rightarrow\mathbf{E}_{n}\rightarrow\mathbf{Z}_{n+1}.\label{eq:SR-trial}
\end{equation}
Intuitively, the trial structure works the following way. Trial $n$
starts with a certain state of conditioning and a sampled stimulus.
Once a stimulus is sampled, a response is computed according to the
state of conditioning. Then, reinforcement follows, which can lead
(with probability $c$) to a new state of conditioning for trial $n+1$
. In more detail, at the beginning of a trial, the state of conditioning
is represented by the random variable $\mathbf{Z}_{n}=\left(z_{1}^{\left(n\right)},\ldots,z_{m}^{\left(n\right)}\right)$.
The vector $\left(z_{1}^{\left(n\right)},\ldots,z_{m}^{\left(n\right)}\right)$
associates to each stimuli $s_{i}\in S$, $i=1,\ldots,m$, where $m=\left|S\right|$
is the cardinality of $S$, a value $z_{i}^{\left(n\right)}$ on trial
$n$. Once a stimulus $\mathbf{S}_{n}=s_{i}$ is sampled with probability
$P\left(\mathbf{S}_{n}=s_{i}|s_{i}\epsilon S\right)=\dfrac{1}{m}$,
its corresponding $z_{i}^{\left(n\right)}$ determines the probability
of responses in $R$ by the probability distribution $K\left(r|z_{i}^{\left(n\right)}\right)$,
i.e. $P\left(a_{1}\leq\mathbf{R}_{n}\leq a_{2}|\mathbf{S}_{n}=s_{i},\mathbf{Z}_{n,i}=z_{i}^{\left(n\right)}\right)=\int_{a_{1}}^{a_{2}}k\left(x|z_{i}^{\left(n\right)}\right)dx$,
where $k\left(x|z_{i}^{\left(n\right)}\right)$ is the probability
density associate to the distribution, and where $\mathbf{Z}_{n,i}$
is the $i$-th component of the vector $\left(z_{1}^{\left(n\right)},\ldots,z_{m}^{\left(n\right)}\right)$.
The probability distribution $K\left(r|z_{i}^{\left(n\right)}\right)$
is the smearing distribution, and it is determined by its variance
and mode $z_{i}^{\left(n\right)}$ . The next step is the reinforcement
$\mathbf{E}_{n}$, which is effective with probability $c$, i.e.
$P\left(\mathbf{Z}_{n+1,i}=y|\mathbf{S}_{n}=s_{i},\mathbf{E}_{n}=y,\mathbf{Z}_{n,i}=z_{i}^{\left(n\right)}\right)=c$
and $P\left(\mathbf{Z}_{n+1,i}=z_{i}^{\left(n\right)}|\mathbf{S}_{n}=s_{i},\mathbf{E}_{n}=y,\mathbf{Z}_{n,i}=z_{i}^{\left(n\right)}\right)=1-c$.
The trial ends with a new (with probability $c$) state of conditioning
$\mathbf{Z}_{n+1}$.

\subsection{Qualitative description of oscillator model\label{sub:Qualitative-description-of}}

Because of its focus on behavioral outcomes, mathematical SR theory
does not provide a clear connection to the internal processing of
information by the brain. To bridge this gap, \citet{suppes_phase-oscillator_2012}
proposed a neural-oscillator based response computation model able
to reproduce the main stochastic features of SR theory, including
the conditional probabilities for a continuum of responses. It is
this model that we describe qualitatively in this subsection. 

In SR theory, a trial is given by the sequence in (\ref{eq:SR-trial}).
In the oscillator model, with the exception of $\mathbf{Z}_{n}$,
each random variable is represented by a corresponding oscillator.
To better understand the model, let us examine each step of an SR
trial in terms of oscillators. 

Let us start with the sampling of a stimulus, $\mathbf{S}_{n}$. At
the beginning of a trial, a (distal) stimulus is presented. Through
the perceptual system processing, it reaches an area of the brain
where it produces a somewhat synchronized firing of a set of neurons.
This set of firing neurons is what we think as the activation of the
sampled stimulus oscillator%
\footnote{We limit our discussion to a single stimulus oscillator, but its generalization
to $N$ stimuli is straightforward.%
}, $O_{s}\left(t\right)$. The oscillator $O_{s}\left(t\right)$ is
a simplified description of the semi-periodic activity of the collection
of (perhaps thousands of) neurons. 

We now turn to the computation of a response, $\mathbf{R}_{n}$. For
simplicity, we first consider two possible responses, $R_{1}$ and
$R_{2}$ (i.e. $R=\left\{ R_{1},R_{2}\right\} $); later we show how
we generalize to any number of to a continuum of responses. Once $O_{s}\left(t\right)$
is activated, its couplings to other neurons may lead to a spreading
of its activation to other sets of neurons, including the response
neurons. Say we can split the neurons activated by $O_{s}\left(t\right)$'s
couplings into two sets, one representing $R_{1}$ and the other $R_{2}$.
Since those neurons were activated, as the stimulus neurons, they
fire in relative synchrony, and can be represented by the oscillators
$O_{r_{1}}\left(t\right)$ and $O_{r_{2}}\left(t\right)$. Of course,
the activation of both $O_{r_{1}}\left(t\right)$ and $O_{r_{2}}\left(t\right)$
is not enough for the computation of a response. To see how a response
is computed in the oscillator model from $O_{r_{1}}\left(t\right)$
and $O_{r_{2}}\left(t\right)$, we need think of what happens if two
oscillators are in-phase or out-of-phase. When two neural oscillators
are in phase, the firing of one adds to the firing of the other, perhaps
reaching a certain response threshold; if they are out of phase, such
threshold may not be reached, as the firings do not add above their
activated intensities. So, a response is chosen depending on the relative
phases of the oscillators. If $O_{s}\left(t\right)$ is in phase with
$O_{r_{1}}\left(t\right)$ and out of phase with $O_{r_{2}}\left(t\right)$,
response $R_{1}$ is selected. Which response is selected depends
on the actual couplings between $O_{s}\left(t\right)$, $O_{r_{1}}\left(t\right)$,
and $O_{r_{2}}\left(t\right)$ (see next subsection for details). 

We can now talk about $\mathbf{E}_{n}$, $\mathbf{Z}_{n}$, and $\mathbf{Z}_{n+1}$.
Since the selected response is determined by the couplings between
neural oscillators, clearly the state of conditioning $\mathbf{Z}_{n}$
depends on those couplings. Therefore, on the oscillator model, saying
that a trial starts with a state of conditioning $\mathbf{Z}_{n}$
is the same as saying that a trial starts with a set of synaptic connections
between the neurons belonging to the ensembles representing the activated
stimulus and responses. But, as the reinforcement $\mathbf{E}_{n}$
may change $\mathbf{Z}_{n}$ into a different state of conditioning,
$\mathbf{Z}_{n+1}$, it follows that during reinforcement the neural
couplings need to change. So, in the oscillator model, such couplings
change when a neural reinforcement oscillator (corresponding to the
activation of an ensemble of neurons representing reinforcement) interacts
with the stimulus and response oscillators, thus driving a Hebb-like
change in synaptic connection, resulting in learning. 

The previous paragraphs summarize qualitatively the model, but there
are three points that need to be further discussed before we proceed
to the quantitative description. First, since neural oscillators correspond
to a tremendous dimensional reduction in the neuronal dynamical system,
we need to justify their use. Second, as it is crucial for the model,
an argument needs to be made as to how oscillators synchronize. Finally,
we should the two-response case, but how can a finite number of oscillators
encode a large number of responses, possibly a continuum of responses? 

Why neural oscillators? First, we note that \citet{suppes_phase-oscillator_2012}
does not propose the use of neural oscillators for any type of modeling.
Instead, they argue that if we are interested in higher cognitive
functions, such as language, it can be showed that they can be represented
on electroencephalogram (EEG) oscillation patterns \citep[see ][Chapter 8, and corresponding references]{suppes_representation_2002}.
But EEG signals are the result of a large assembly of neurons firing
synchronously or quasi-synchronously \citep{nunez_electric_2006}.
Thus, we can think of a word as represented in the brain by a set
of coupled neurons. When this word is activated by an external stimulus,
this set of neurons fire coherently. Such an ensemble of synchronizing
neurons could, in first approximation, be represented by a periodic
function $F_{T}(t)$, where $T$ is the period of the function%
\footnote{We remark that neurons in this ensemble do not need to fire at the
same time, but only that their frequencies synchronize. They may still
be in synchrony but out of phase. %
}. So, for \citet{suppes_phase-oscillator_2012} a neural oscillator
is simply a set of firing neurons represented by such periodic functions,
and a good candidate as a unit of processing of information in the
brain. 

It must be emphasized that representing a collection of neurons by
a periodic function, as we did above, does not mean that all neurons
fire simultaneously. It just means that the firing frequencies of
individual neurons become very close to each other. But neurons with
the same frequency do not need to fire at the same times. In other
words, they may synchronize out of phase. But, more importantly, once
they synchronize, the initial dynamical system with thousands of degrees
of freedom become describable by a single dynamical variable, the
periodic function $F_{T}(t)$. It is this dramatic dimensional reduction
that allows us to produce a model whose behavior we can understand
in certain specific situations. 

We now move to the question of how neural oscillators synchronize.
Let us first look at the qualitative behavior of individual neurons
in two ensembles, $A$ and $B$, each with a large number of neurons
firing coherently. Let $n_{A}$ be a single neuron in ensemble $A$
firing with period $T_{A}$, and let $n_{B}$ be another neuron in
$B$ firing with period $T_{B}$. Let $t'$ be a particular time such
that $n_{B}$ would fire if $A$ were not activated, i.e. the time
it would fire because of its natural periodicity $T_{B}$. If $n_{A}$
and $n_{B}$ are synaptically coupled, a firing of $n_{A}$ at a time
$t\lessapprox t'$ would anticipate $n_{B}$'s firing to a time $t''$
closer to $t$, thus approaching the timing of firing for both neurons%
\footnote{This is true for excitatory synapses. As we shall see later on this
paper, inhibitory synapses will play a different yet important role.%
}. If more neurons from ensemble $A$ are coupled to $n_{B}$, the
stronger the effect of anticipating its firing time. Furthermore,
since $n_{B}$ is also coupled to neurons in $A$, its firing has
the same effect on $A$ (though this effect does not need to be symmetric).
So, the couplings between $A$ and $B$ lead both periods $T_{A}$
and $T_{B}$ to approach each other, i.e., $A$ and $B$ synchronize.
In fact, it is possible to prove mathematically that, under the right
conditions, if the number of neurons is large enough, the sum of the
several weak synaptic interactions can cause a strong effect, making
all neurons fire close together \citep{izhikevich_dynamical_2007}.
In other words, even when weakly coupled, two neural ensembles represented
by oscillators may synchronize. 

We end this subsection with the representation of a continuum of responses
in terms of neural oscillators. Continuum of responses show up in
some behavioral experiments. For example, a possible task could be
to predict the position of an airplane on a radar screen (represented
by a flashing dot appearing periodically) after observing its position
for several trials. Since the radar position can be anywhere in the
screen, this cannot be represented by a discrete set of responses. 

So, how could we represent a continuum of responses with neural oscillators?
With a discrete number of responses, what we did was to create a representation
for each response in terms of a distinct oscillator (in our example
above, $O_{r_{1}}\left(t\right)$ and $O_{r_{2}}\left(t\right)$).
But this approach would lead to an uncountable number of neural oscillators
for the continuum case, clearly a physically unreasonable assumption.
But there is another alternative. When we used oscillators for the
two responses, we thought about $O_{r_{1}}\left(t\right)$ and $O_{r_{2}}\left(t\right)$
as being either in phase or out of phase with $O_{s}\left(t\right)$.
However, they do not need to be in perfect phase. In the next section,
we show that by using specific differences of phase between $O_{r_{1}}\left(t\right)$
and $O_{r_{2}}\left(t\right)$ and $O_{s}\left(t\right)$, we can
control the amount of ``intensity'' of a neural activation on each
oscillator. This relative intensity between $O_{r_{1}}\left(t\right)$
and $O_{r_{2}}\left(t\right)$ can be used as a measure of how close
a response is to either $R_{1}$ or $R_{2}$; in other words, we can
think of the response as being between $R_{1}$ and $R_{2}$, which
would correspond to a continuum. So, by thinking of $R_{1}$ and $R_{2}$
as extremes in a possible range of responses in a continuum interval,
the relative intensities of $O_{r_{1}}\left(t\right)$ and $O_{r_{2}}\left(t\right)$,
given by their phase relation with $O_{s}\left(t\right)$, could code
any response in between.

\subsection{Quantitative description of the oscillator model\label{sub:Quantitative-description-of}}

We now turn to a more rigorous mathematical description. Our underlying
assumption is that we have two neural ensembles $A$ and $B$ described
by periodic functions $F_{T_{A}}\left(t\right)$ and $F_{T_{B}}\left(t\right)$
with periods $T_{A}$ and $T_{B}$, respectively. Without loss of
generality in our argument, we assume that $F_{T_{A}}\left(t\right)$
and $F_{T_{B}}\left(t\right)$ have the same shape, i.e., there exists
a constant $c$ such that $F_{T_{A}}\left(t\right)=F_{T_{B}}\left(tc\right)$.
A simple way to describe the synchronization of $A$ and $B$ is to
rewrite the time argument in the functions, rewriting them as $F_{T_{A}}\left(t\alpha_{A}\right)$
and $F_{T_{B}}\left(t\alpha_{B}\right)$. Clearly, if no couplings
exist between $A$ and $B$, their dynamics is not modified and $\alpha_{A}=\alpha_{B}=1$.
However, when the ensembles are couples, $\alpha_{A}$ and $\alpha_{B}$
are a function of time, their synchronization means that they evolve
such that $F_{T_{A}}\left(t\alpha_{A}\left(t\right)\right)=F_{T_{B}}\left(t\alpha_{B}\left(t\right)\right)$.
Thus, when we are studying the dynamics of coupled neural oscillators,
it suffices to study the dynamics of the arguments of $F_{T_{A}}$
and $F_{T_{B}}$ above, which we call their \emph{phases}. 

To illustrate this point, let us imagine a simple case where when
there is no interaction $A$ follows a harmonic oscillator with angular
frequency $\omega_{A}=2\pi/T_{A}$. Then $F_{T_{A}}\left(t\right)=A\cos\left(\omega_{A}t+\delta_{A}\right)$,
where and $\delta_{A}$ is a constant, and expressions hold for $F_{T_{B}}\left(t\right)$.
Following the above paragraph, we rewrite 
\[
F_{T_{A}}\left(t\right)=A\cos\left(\varphi_{A}\left(t\right)\right),
\]
where 
\begin{equation}
\varphi_{A}\left(t\right)=\omega_{A}t+\delta_{A}\label{eq:phi-solution-no-interaction}
\end{equation}
is the phase. Let us now focus on the phase dynamics. 

For a set of coupled phase oscillators, \citet{kuramoto_chemical_1984}
proposed a simple set of dynamical equations for $\varphi_{A}\left(t\right)$.
To better understand them, let us assume that in the absence of interactions
$\varphi_{A}$ evolves according to (\ref{eq:phi-solution-no-interaction}).
Thus, the differential equation describing $\varphi_{A}$ would be
\begin{equation}
\frac{d\varphi_{A}\left(t\right)}{dt}=\omega_{A},\label{eq:kura-no-interaction-A}
\end{equation}
and for $\varphi_{B}$,
\begin{equation}
\frac{d\varphi_{B}\left(t\right)}{dt}=\omega_{B}.\label{eq:kura-no-interaction-B}
\end{equation}
In the presence of a synchronizing interaction, Kuramoto proposed
that equations (\ref{eq:kura-no-interaction-A}) and (\ref{eq:kura-no-interaction-B})
should be replaced by 
\begin{equation}
\frac{d\varphi_{A}\left(t\right)}{dt}=\omega_{A}-k_{AB}\sin\left(\varphi_{A}\left(t\right)-\varphi_{B}\left(t\right)\right),\label{eq:kura-interaction-A}
\end{equation}
and for $\varphi_{B}$,
\begin{equation}
\frac{d\varphi_{B}\left(t\right)}{dt}=\omega_{B}-k_{BA}\sin\left(\varphi_{B}\left(t\right)-\varphi_{A}\left(t\right)\right).\label{eq:kura-interaction-B}
\end{equation}
It is easy to see how equations (\ref{eq:kura-interaction-A}) and
(\ref{eq:kura-interaction-B}) work. When the phases $\varphi_{A}$
and $\varphi_{B}$ are close to each other, we can linearize the sine
term, and rewrite equations (\ref{eq:kura-interaction-A}) and (\ref{eq:kura-interaction-B})
as 
\begin{equation}
\frac{d\varphi_{A}\left(t\right)}{dt}\cong\omega_{A}t-k_{AB}\left(\varphi_{A}\left(t\right)-\varphi_{B}\left(t\right)\right),\label{eq:kura-interaction-A-linear}
\end{equation}
and for $\varphi_{B}$,
\begin{equation}
\frac{d\varphi_{B}\left(t\right)}{dt}\cong\omega_{B}t-k_{BA}\left(\varphi_{B}\left(t\right)-\varphi_{A}\left(t\right)\right).\label{eq:kura-interaction-B-linear}
\end{equation}
The effect of the added term in equation (\ref{eq:kura-interaction-A-linear})
is to make the instantaneous frequency increase above its natural
frequency $\omega_{A}$ if $\varphi_{B}$ is greater than $\varphi_{A}$,
and decrease otherwise. In other words, it pushes the phases $\varphi_{A}$
and $\varphi_{B}$ towards synchronization. We should emphasize that
Kuramoto's equations were proposed for a large number of oscillators,
and his goal was to find exact solutions for such large systems. However,
as we described above, his equations also make sense for systems with
small number of oscillators (for applications of Kuramoto's equations
to systems with small number of oscillators, in addition to \citet{suppes_phase-oscillator_2012},
see also \citet{billock_sensory_2005,billock_honor_2011,seliger_plasticity_2002,trevisan_dynamics_2005}
and references therein). For a set of $N$ coupled phase oscillators,
Kuramoto's equations are 
\begin{equation}
\frac{d\varphi_{i}\left(t\right)}{dt}=\omega_{i}-\sum_{\substack{j=1\\
j\neq i
}
}^{N}k_{ij}\sin\left(\varphi_{i}\left(t\right)-\varphi_{j}\left(t\right)\right).\label{eq:kura-n-osc}
\end{equation}
But independent of whether we have large or small number of oscillators
or not, it is possible to prove that for oscillating dynamical systems
near a bifurcation (which is the case for many neuronal models), Kuramoto's
equations present a first order approximation for the synchronizing
coupled dynamics \citep{izhikevich_dynamical_2007}. 

In the model proposed in \citep{suppes_phase-oscillator_2012}, each
stimulus in the set $S$ of stimuli corresponds to an oscillator,
i.e. there are $m$ stimulus neural oscillators, $\left\{ s_{i}\left(t\right)\right\} $,
$i=1,\ldots,N$. Here we use the notation $s_{i}\left(t\right)$ to
distinguish between the oscillator and the actual stimulus $s_{i}\in S$.
Once a distal stimulus is presented, the processing of the proximal
stimulus leads, after sensory processing, to the activation of neurons
in the brain corresponding to a neural oscillator representation of
such stimulus. Once a stimulus oscillator is activated by starting
to fire synchronously, a set of response oscillators, $r_{1}\left(t\right)$
and $r_{2}\left(t\right)$, connected to the active stimulus oscillator
via the couplings $k_{s_{i},r_{j}}$ and $k_{r_{1},r_{2}}$ lead to
the synchronization of the stimulus and response oscillators. The
couplings $k_{s_{i},r_{j}}$ and $k_{r_{1},r_{2}}$ correspond to
the state of conditioning in SR-theory. Depending on the details of
the synchronization dynamics, as explained below, a response is selected
(this being the equivalent of sampling $\mathbf{X}_{n}$). Finally,
during reinforcement, a reinforcement oscillator $e_{y}\left(t\right)$
is activated, and its couplings with $s_{i}\left(t\right)$, $r_{1}\left(t\right)$,
and $r_{2}\left(t\right)$, together with $k_{s_{i},r_{j}}$ and $k_{r_{1},r_{2}}$,
leads to a dynamics that allow for changes in $k_{s_{i},r_{j}}$ and
$k_{r_{1},r_{2}}$, which accounts for the last step in the SR trial
(\ref{eq:SR-trial}), with a new state of conditioning. 

Before we detail the dynamics of the oscillator model, it is important
to discuss how $s_{i}\left(t\right)$, $r_{1}\left(t\right)$, and
$r_{2}\left(t\right)$ can encode a response through their couplings,
and how can we interpret such couplings. Let us write the case where
$s_{i}$ is sampled, and let us assume that once the dynamics is acting,
all oscillators synchronize, acquiring the same frequency (though
perhaps with a phase difference). Then, 
\begin{eqnarray}
s_{i}(t) & =A\cos\left(\varphi_{s_{i}}(t)\right)= & A\cos\left(\omega_{0}t\right),\label{eq:oscillation-s}\\
r_{1}(t) & =A\cos\left(\varphi_{r_{1}}(t)\right)= & A\cos\left(\omega_{0}t+\delta\phi_{1}\right),\label{eq:oscillation-1}\\
r_{2}(t) & =A\cos\left(\varphi_{r_{2}}(t)\right)= & A\cos\left(\omega_{0}t+\delta\phi_{2}\right),\label{eq:oscillation-2}
\end{eqnarray}
where $s(t)$, $r_{1}(t)$, and $r_{2}(t)$ represent harmonic oscillations,
$\varphi_{s}(t)$, $\varphi_{r_{1}}(t)$, and $\varphi_{r_{2}}(t)$
their phases, $\delta\phi_{1}$ and $\delta\phi_{2}$ are constants,
and $A$ their amplitude, assumed to be the same. As argued in \citep{suppes_phase-oscillator_2012},
since neural oscillators have a wave-like behavior \citep{nunez_electric_2006},
their dynamics satisfy the principle of superposition, thus making
oscillators prone to interference effects. As such, the mean intensity
in time gives us a measure of the excitation of the neurons in the
neural oscillators. For the superposition of $s(t)$ and $r_{1}(t)$,
\begin{eqnarray*}
I_{1} & = & \left\langle \left(s_{i}(t)+r_{1}(t)\right)^{2}\right\rangle _{t}\\
 & = & \left\langle s(t)^{2}\right\rangle _{t}+\left\langle r_{1}(t)^{2}\right\rangle _{t}+\left\langle 2s(t)r_{1}(t)\right\rangle _{t},
\end{eqnarray*}
where 
\[
\left\langle f\left(t\right)\right\rangle _{t}=\lim_{T\rightarrow\infty}\frac{1}{T}\int_{0}^{T}f\left(t\right)dt
\]
is the time average. It is easy to compute that 
\[
I_{1}=A^{2}\left(1+\cos\left(\delta\phi_{1}\right)\right),
\]
and 
\begin{eqnarray*}
I_{2} & = & A^{2}\left(1+\cos\left(\delta\phi_{2}\right)\right).
\end{eqnarray*}
From the above equations, the maximum intensity for each superposition
is $2A^{2}$, and the minimum is zero. Thus, the maximum difference
between $I_{1}$ and $I_{2}$ happens when their relative phases are
$\pi$. We also expect a maximum contrast between $I_{1}$ and $I_{2}$
when a response is not in between the responses represented by the
oscillators $r_{1}\left(t\right)$ and $r_{2}\left(t\right)$. For
in between responses, we should expect less contrast, with the minimum
contrast happening when the response lies on the mid-point of the
continuum between the responses associated to $r_{1}(t)$ and $r_{2}(t)$.
The ideal balance of responses happen if we impose
\begin{equation}
\delta\phi_{1}=\delta\phi_{2}+\pi\equiv\delta\phi,\label{eq:ideal-phase-diff}
\end{equation}
which results in 
\begin{equation}
I_{1}=A^{2}\left(1+\cos\left(\delta\phi\right)\right),\label{eq:phase-1}
\end{equation}
 and
\begin{equation}
I_{2}=A^{2}\left(1-\cos\left(\delta\phi\right)\right).\label{eq:phase-2}
\end{equation}
From equations (\ref{eq:phase-1}) and (\ref{eq:phase-2}), let $x\in[-1,1]$
be the normalized difference in intensities between $r_{1}$ and $r_{2}$,
i.e. 
\begin{eqnarray}
x & \equiv & \frac{I_{1}-I_{2}}{I_{1}+I_{2}}=\cos\left(\delta\varphi\right),\label{eq:angle-reinforcement-b}
\end{eqnarray}
$0\leq\delta\varphi\leq\pi$. The parameter $x$ codifies the dispute
between response oscillators $r_{1}\left(t\right)$ and $r_{2}\left(t\right)$.
So, we can use arbitrary phase differences between oscillators to
code for a continuum of responses between $-1$ and $1$.

As we mentioned above, the dynamics of the phase oscillators leading
to (\ref{eq:oscillation-s})--(\ref{eq:oscillation-2}) are modeled
here by Kuramoto equations (\ref{eq:kura-n-osc}). However, as we
also mentioned, (\ref{eq:kura-n-osc}) leads to the synchronization
of all oscillators with the same phase, which is not what we need
to code responses according to equation (\ref{eq:angle-reinforcement-b}).
Naturally, Kuramoto's equations need to be modified to encode the
phase differences $\delta\varphi$ in (\ref{eq:oscillation-s})--(\ref{eq:oscillation-2}).
This can be accomplished by adding a term inside the sine, resulting
in 
\begin{equation}
\frac{d\varphi_{i}\left(t\right)}{dt}=\omega_{i}-\sum_{\substack{j=1\\
j\neq i
}
}^{N}k_{ij}\sin\left(\varphi_{i}\left(t\right)-\varphi_{j}\left(t\right)+\delta_{ij}\right).\label{eq:Kuramoto-phase-differences}
\end{equation}
The main problem with such change is how to interpret it. Equation
(\ref{eq:kura-n-osc}) had a clear interpretation: the firing of oscillator
neurons coupled through excitatory synapses displaced the phase of
the other oscillator into synchrony. But this interpretation does
not make sense for (\ref{eq:Kuramoto-phase-differences}). Why would
oscillators be brought close to each other, but be kept at a phase
distance of $\delta_{ij}$? 

To understand the origin of $\delta_{ij}$, let us rewrite (\ref{eq:Kuramoto-phase-differences})
as\textbf{ 
\begin{eqnarray}
\frac{d\varphi_{i}}{dt} & = & \omega_{i}-\sum_{\substack{j=1\\
j\neq i
}
}^{N}k_{ij}\cos\left(\delta_{ij}\right)\sin\left(\varphi_{i}-\varphi_{j}\right)\nonumber \\
 &  & -\sum_{\substack{j=1\\
j\neq i
}
}^{N}k_{ij}\sin\left(\delta_{ij}\right)\cos\left(\varphi_{i}-\varphi_{j}\right).\label{eq:kura-phase-rewritten}
\end{eqnarray}
}Since the terms involving the phase differences $\delta_{ij}$ are
constant, we can write (\ref{eq:kura-phase-rewritten}) as\textbf{
\begin{equation}
\frac{d\varphi_{i}}{dt}=\omega_{i}-\sum_{\substack{j=1\\
j\neq i
}
}^{N}\left[k_{ij}^{E}\sin\left(\varphi_{i}-\varphi_{j}\right)+k_{ij}^{I}\cos\left(\varphi_{i}-\varphi_{j}\right)\right],\label{eq:kuramoto-equations-inhibition-excitation}
\end{equation}
}where $k_{ij}^{E}\equiv k_{ij}\cos\left(\delta_{ij}\right)$ and
$k_{ij}^{I}\equiv k_{ij}\sin\left(\delta_{ij}\right)$. Equation (\ref{eq:Kuramoto-phase-differences})
now has a clear interpretation: $k_{ij}^{E}$ makes oscillators $\varphi_{i}\left(t\right)$
and $\varphi_{j}\left(t\right)$ approach each other, $k_{ij}^{E}$
makes them move further apart. Thus, $k_{ij}^{E}$ corresponds to
excitatory couplings between neurons and $k_{ij}^{I}$ to inhibitory
couplings. In other words, we give meaning to equations (\ref{eq:Kuramoto-phase-differences})
by rethinking their couplings in terms of excitatory and inhibitory
neuronal connections. 

To summarize the response process, once a stimulus oscillator $s_{i}\left(t\right)$
is activated at time $t_{s,n}$ on trial $n$, the response oscillators
are also activated. Because of the excitatory and inhibitory couplings
between stimulus and response oscillators, after a certain time their
dynamics lead to their synchronization, but with phase differences
given by (\ref{eq:ideal-phase-diff}). Here we point out that, due
to stochastic variations of biological origin, the initial conditions
$s_{i}\left(t_{s,n}\right)$, $r_{1}\left(t_{s,n}\right)$, and $r_{2}\left(t_{s,n}\right)$
vary according to the distribution
\begin{equation}
f\left(\varphi_{i}\right)=\frac{1}{\sigma_{\varphi}\sqrt{2\pi}}\exp\left(-\frac{\varphi_{i}}{2\sigma_{\varphi}^{2}}\right).\label{eq:phase-density}
\end{equation}
Therefore, the phase differences at the time of response, $t_{x,n}$,
may not be exactly the ones given in (\ref{eq:ideal-phase-diff}),
which gives an underlying explanation for a component of the smearing
distribution%
\footnote{We note that we here only model the brain computation of a stimulus
and response, and not the representation of the distal stimulus into
brain oscillators and the representation of responses into actual
muscle movement or sound production. Of course, those extra steps
between the stimulus and the response will lead to further smearing
of the actual observed distribution, and a detailed brain computation
of $K\left(r|z_{i}^{\left(n\right)}\right)$ should include such factors.
The interested reader should refer to \citet{suppes_phase-oscillator_2012}
for details about the used parameters and the fitting of the model
to experimental data.%
}. Interference between oscillators, determined by their phase differences,
leads to a relative intensity that codes a continuum of responses
according to $x$ in equation (\ref{eq:angle-reinforcement-b}). The
variable $x$ is the response computed by the oscillators. 

We now turn to learning. Here we will only present the main features
of learning relevant to this paper. The actual details are numerous,
and we refer the to \citep{suppes_phase-oscillator_2012}. For learning
to happen, the couplings $k_{ij}^{E}$ and $k_{ij}^{I}$ need to change
during reinforcement. Since reinforcement must have a brain representation
in terms of synchronously firing neurons, during reinforcement a neural
oscillator $\varphi_{y}\left(t\right)$ gets activated. We assume
that $\varphi_{y}\left(t\right)$ is a fixed representation in the
brain, with frequencies that are independent of the stimulus and response
oscillators, and therefore can be represented by 
\[
\varphi_{y}\left(t\right)=\omega_{e}t.
\]
Furthermore, since $\varphi_{y}$ is coding a response $y$ being
reinforced, from (\ref{eq:angle-reinforcement-b}) the phase of the
oscillators must be related to $y$ via 
\[
\delta\varphi=\arccos y.
\]
During the reinforcement period, the (active) stimulus and response
oscillators evolve according to the same equations as before, but
now with an extra term due to the reinforcement oscillator, according
to 
\begin{eqnarray}
\frac{d\varphi_{i}}{dt} & = & \omega_{i}-\sum_{s_{i},r_{1},r_{2}}k_{ij}^{E}\sin\left(\varphi_{i}-\varphi_{j}\right)\nonumber \\
 &  & -\sum_{s_{i},r_{1},r_{2}}k^{E}\sin\left(\varphi_{i}-\varphi_{j}\right)\nonumber \\
 &  & -K_{0}\sin\left(\varphi_{s_{j}}-\omega_{e}t+\Delta_{i}\right),\label{eq:learning-phase}
\end{eqnarray}
where $\Delta_{ij}$ reflects the reinforced response $y$, 
\[
\Delta_{i}=-\left(\delta_{i,r_{1}}+\delta_{i,r_{2}}\right)\arccos y+\delta_{i,r_{2}}\pi,
\]
and where $\delta_{i,j}$ is Kronecker's delta. During reinforcement,
couplings also change according to a Hebbian rule represented by the
equations \citep{seliger_plasticity_2002}
\begin{eqnarray}
\frac{dk_{ij}^{E}}{dt} & = & \epsilon\left(K_{0}\right)\left[\alpha\cos\left(\varphi_{i}-\varphi_{j}\right)-k_{ij}^{E}\right],\label{eq:learning-asym-excitatory}
\end{eqnarray}
and 
\begin{eqnarray}
\frac{dk_{ij}^{I}}{dt} & = & \epsilon\left(K_{0}\right)\left[\alpha\sin\left(\varphi_{i}-\varphi_{j}\right)-k_{ij}^{I}\right],\label{eq:learning-asym-inhibitory}
\end{eqnarray}
where 
\begin{equation}
\epsilon\left(K_{0}\right)=\left\{ \begin{array}{c}
0\mbox{ if }K_{0}<K'\\
\epsilon_{0}\mbox{ otherwise},
\end{array}\right.\label{eq:epsilon}
\end{equation}
where $\epsilon_{0}\ll\omega_{0}$, $\alpha$ and $K_{0}$ are constant
during $\Delta t_{e}$, and $K'$ is a threshold constant throughout
trials \citep{hoppensteadt_synaptic_1996-1,hoppensteadt_synaptic_1996}.
We also assume that that there is a normal probability distribution
governing the coupling strength $K_{0}$ between the reinforcement
and the other active oscillators. It has mean $\overline{K}_{0}$
and standard deviation $\sigma_{K_{0}}$. Its density function is:
\begin{equation}
f\left(K_{0}\right)=\frac{1}{\sigma_{K_{0}}\sqrt{2\pi}}\exp\left\{ -\frac{1}{2\sigma_{K_{0}}^{2}}\left(K_{0}-\overline{K}_{0}\right)^{2}\right\} .\label{eq:K0-density}
\end{equation}
So, when a reinforcement oscillator is activated, the system evolves
under the coupled set of differential equations (\ref{eq:learning-phase}),
(\ref{eq:learning-asym-excitatory}), and (\ref{eq:learning-asym-inhibitory}).

To summarize, we described a model of SR theory in terms of neural
oscillators. In such a model, stimulus and responses are represented
by collections of neurons firing in synchrony. Such collections of
neurons, when coupled, synchronize to each other. The relative phases
of synchronization lead to interference of sets of neurons, and such
interference results in the coding of a continuum of responses. We
left our many of the details of the stochastic processes involved
in this model, as they are not necessary for this paper, but details
can be found in reference \citep{suppes_phase-oscillator_2012}.

\section{Quantum-like neural oscillator effects}

In this section we discuss possible quantum-like effects in the brain
and how they can be understood in terms of oscillators. We start by
discussing what we believe are quantum-like effects in the brain.
We then describe how quantum-like effects show up in decision-making
experiments. Finally, we show that neural oscillators can reproduce
such effects.

\subsection{What are quantum-like effects?\label{sec:Quantum-like-effects}}

As we mentioned earlier, quantum mechanics is one of the most successful
theories in history. However, such success did not come without a
price, as quantum mechanics presents a view of the world that most
people would find disturbing, to say the least. This was certainly
the view of many of the founders of quantum mechanics, such as Plank,
Einstein, and Pauli. At the core of their concerns were three characteristics
of quantum mechanics: nondeterminism, contextuality, and nonlocality.
Let us examine each one of those characteristics, and see how they
can be relevant to brain processes. 

We start with nondeterminism. Intuitively, a dynamical system is deterministic
if the current state of the system completely determines the future
state of such system, and it is nondeterministic otherwise. Quantum
mechanics is nondeterministic because a quantum state only gives the
probabilities for the values of certain observables during a measurement
process. Of course, nondeterminism is not unique to quantum mechanics.
For practical purposes, there are many nondeterministic processes,
such as the famous Brownian motion, or the tossing of a coin. But
when we toss a coin or throw a die, physicists believe that the newtonian
dynamics would allow for the computation of the final state of the
die or coin, if we were to know exactly the initial conditions, i.e.,
the state of the system \citep{ford_how_1983,vulovic_randomness_1986}.
In other words, before quantum mechanics physicists used to believe
that the underlying dynamics for nondeterministic processes was actually
deterministic; we just could not distinguish the deterministic from
the nondeterministic because we could not know enough details about
the system. But whether we can distinguish between a deterministic
and stochastic dynamics is not the important issue (in fact sometimes
we can't; see \citet{weingartner_photons_1996} for detail). The main
point is that for quantum theory the underlying dynamics is essentially
nondeterministic. 

Nondeterministic stochastic processes are often used in cognitive
modeling \citep{busemeyer_cognitive_2010}. For example, the stochastic
SR theory mentioned above is nondeterministic, as conditioning happens
probabilistically. Yet, as in the coin tossing example, there is no
reason to believe that the underlying dynamics must be nondeterministic,
as its stochasticity may come from our lack of detailed knowledge
about the brain or its biological noises \citep{josic_coherent_2009}.
So, it should be clear that when we talk about quantum-like brain
processes, nondeterminism should not play a prominent role. Insofar
as nondeterminism is concerned, there is no need to add further quantum-mechanical
mathematical or conceptual machinery to account for it in the brain. 

Now let us examine nonlocality. To discuss nonlocality, we should
first discuss causality, albeit briefly (interested readers are referred
to \citet{suppes_probabilistic_1970}). Let $A$ and $B$ be two events,
and let $\mathbf{A}$ and $\mathbf{B}$ be $\pm1$-random variables
corresponding to whether $A$ and\textbf{ $B$ }occurred ($+1$ if
occurred, $-1$ otherwise). We say that $C$ is a\emph{ cause of $A$
}if $P(\mathbf{A}=1|\mathbf{C}=1)>P\left(\mathbf{A}=1\right)$ and
there not exist a spurious cause (represented by a hidden variable
$\mathbf{D}$) that could account for the correlations. Now, it is
possible to prove that for some quantum systems, there are space-like
events $A$ and $A'$ such that $A$ is a the cause of $A'$. What
is key in the previous statement is that $A$ and $A'$ are space-like
separated events; in other words, whatever mechanism is making $A'$
influence $A$ must be superluminal%
\footnote{We are summarizing a very complex and subtle discussion in a few sentences.
The reader interested in more details and in a mathematically rigorous
treatment is directed to \citet{suppes_representation_2002} and references
therein. %
}. This, it may be argued, is the main puzzling characteristic of quantum
processes: to understand them we need to accept superluminal causation%
\footnote{Or abandon realism, in the sense that the act of measurement of an
observer (a free choice of the experimenter) determines the existence
of the quantity being measured (though realist theories may also have
other problems; see \citet{de_barros_realism_2007}).%
}. However, we believe that quantum-like effects in the brain are not
the outcome of superluminal causation. The brain is small, of the
order of tens of centimeters, and an electromagnetic field would only
need of the order of $10^{-10}$ seconds to travel the whole brain.
Given that brain processes are orders of magnitude slower than $10^{-10}$
seconds, no phenomena in the brain could not be accounted for non-superluminal
causality. In other words, we doubt it would be possible to create
an experiment where nonlocal effects in the brain would be detected
in a superluminal way.

Here we need to differentiate our use of nonlocality from what is
sometimes found in the literature on quantum cognition \citep{nelson_spreading_2003}.
Sometimes, authors consider a violation of some form of Bell's inequalities
\citep{bell_einstein-podolsky-rosen_1964,bell_problem_1966} as evidence
of nonlocality. However, from our discussion, the correlated quantities
must be measured such that no superluminal signal explaining the correlations
is possible, otherwise a violation of Bell's inequalities would only
imply a contextuality \citep{suppes_collection_1996}. Classical fields,
such as the electromagnetic fields used by Khrennikov, also exhibit
contextuality, violating Bell's inequalities and not allowing for
certain configurations a joint probability distribution \citep{suppes_proposed_1996}.
Here we are using nonlocality in the strict sense of ``spooky''
action-at-a-distance correlations that cannot be explained without
using superluminal interactions. 

This brings us to the last quantum-like issue we wish to discuss:
contextuality. Early on, the wave description of a particle led physicists
to realize, through Fourier's theorem, that it was impossible to simultaneously
assign values of momentum and position to it. This developed into
the concept of complementarity. Without trying to be too technical,
we can say that two observables are complementary when they cannot
be simultaneously observed, i.e., when the corresponding hermitian
operators $\hat{O}_{1}$ and $\hat{O}_{2}$ for such properties does
not commute ($[\hat{O}_{1},\hat{O}_{2}]\neq0$). But the question
is not whether we can measure $\hat{O}_{1}$ and $\hat{O}_{2}$ simultaneously,
but whether we can assign those properties values even when we cannot
measure them. In other words, can we generate a data table that fills
out the values of $\hat{O}_{1}$ and $\hat{O}_{2}$ for every trial?
If the number of observables is large enough, the answer to that question
was given in the negative first by \citet{bell_problem_1966}, who
showed that any attempts to fill out such tables would result in values
inconsistent with those predicted by quantum mechanics, and later
by \citet{kochen_problem_1975}. In other words, there are quantum
systems whose observables cannot be assigned values when we do not
measure them, and therefore we cannot provide a consistent joint probability
distribution for such observables \citep{de_barros_inequalities_2000,de_barros_probabilistic_2001,de_barros_probabilistic_2010,de_barros_comments_2011}%
\footnote{Though with the redefinition of what constitutes a particle, local
models reproducing many of the characteristics of quantum mechanics
non-locality are possible. See, for example, \citet{suppes_random-walk_1994,suppes_diffraction_1994,suppes_violation_1996,suppes_particle_1996}
for a particular approach.%
}. 

This impossibility of defining a joint probability distribution is
related to the change of values of the random variables when contexts
change. For example, a random variable $\mathbf{O}_{1}$ representing
$\hat{O}_{1}$ the outcomes of an experiment may not be the same random
variable when $\hat{O}_{2}$ is being measured. Such random variables
change with the context. Contextuality is not new in physics, nor
in psychology. It happens in classical systems, such as with the classical
electromagnetic field \citep{de_barros_quantum_2009}. However, we
claim it is contextuality, with its associated impossibility of defining
joint probability distributions, that leads to quantum-like events
in brain processing. 

At this point, it is important to raise the issue of how much of the
quantum mechanical apparatus is needed to describe brain processes.
Quantum mechanics brings to the table more than just contextuality,
nonlocality, and nondeterminism. It also brings an additional mathematical
structure, formalized by the algebra of observables on a Hilbert space,
initially advocated by \citet{von_neumann_mathematical_1996}. To
see this, let us examine the following case, analyzed in details by
\citet{suppes_when_1981,suppes_collection_1996}. Let $\mathbf{X}$,
$\mathbf{Y}$, and $\mathbf{Z}$ be three $\pm1$-valued random variables
with observed pairwise joint expectations given by 
\[
E\left(\mathbf{XY}\right)=E\left(\mathbf{XZ}\right)=E\left(\mathbf{YZ}\right)=-\epsilon,
\]
where $\epsilon>1/3$. \citet{suppes_when_1981} showed that such
expectations are not consistent with the existence of a joint probability
distribution for $\mathbf{X}$, $\mathbf{Y}$, and $\mathbf{Z}$ if
\[
-1\leq E\left(\mathbf{XY}\right)+E\left(\mathbf{XZ}\right)+E\left(\mathbf{YZ}\right)\leq1+2\min\left\{ E\left(\mathbf{XY}\right),E\left(\mathbf{XZ}\right),E\left(\mathbf{YZ}\right)\right\} .
\]
Thus, if they represent the outcomes of local measurements, they are
contextual. If, on the other hand, they each represent space-like
separated measurement events, they are non-local. And each random
variable is indistinguishable from a tossed coin, and are therefore
nondeterministic%
\footnote{The distinction between nondeterministic and deterministic is a complicated
and subtle one, c.f. \citet{weingartner_photons_1996}, and we refrain
from it in this paper. The interested reader should also refer to
\citet{werndl_are_2009}.%
}. However, $\mathbf{X}$, $\mathbf{Y}$, and $\mathbf{Z}$ cannot
represent quantum mechanical observables, as, since they all commute,
we could in principle simultaneously observe $\mathbf{X}$, $\mathbf{Y}$,
and $\mathbf{Z}$, which is a sufficient condition for the existence
of a joint probability distribution%
\footnote{This comes from the fact that, since they commute with each other,
there exists a set of orthogonal base vectors in the Hilbert space
representation such that the observables corresponding to $\mathbf{X}$,
$\mathbf{Y}$, and $\mathbf{Z}$, namely $\hat{X}$, $\hat{Y}$, and
$\hat{Z}$, are all diagonal in this base. %
}. But we could, on the other hand, imagine a social-science or psychology
example (albeit probably a contrived one) where correlations such
as the ones exhibited above could appear. Furthermore, as shown by
\citet{de_barros_joint_2012}, such random variable correlations can
be obtained from the same neural oscillator model from section \ref{sec:SR-theory-neural-oscillators}. 

In this section we discussed the different ways in which quantum-like
processes may show up in the brain. We argued that among them, the
most probable is contextuality. Contextuality may appear in a system
for several reasons. For example, certain systems evolve according
to a dynamic dependent on boundary conditions, and those may change
with the context. Additionally, a complex system may have different
parts that can interfere, changing their behavior due to alterations
in the conditions of other parts. The latter possibility is mainly
what \citet{khrennikov_quantum-like_2011} explored in his field model.
In the next section, we will show that the complex interaction of
neural oscillators are another plausible explanation for quantum-like
behavior in the brain.

\subsection{Quantum-like effects in decision-making}

The experimental violation of Savage's sure-thing principle (STP)
is considered an example of a quantum-like decision making, so, let
us start with a description of the STP. Savage presents the following
example. 
\begin{quote}
``A businessman contemplates buying a certain piece of property.
He considers the outcome of the next presidential election relevant
to the attractiveness of the purchase. So, to clarify the matter for
himself, he asks whether he should buy if he knew that the Republican
candidate were going to win, and decides that he would do so. Similarly,
he considers whether he would buy if he new that the Democratic candidate
were going to win, and again finds that he would do so. Seeing that
he would buy in either event, he decides that he should buy, even
though he does not know which event obtains, or will obtain, as we
would ordinarily say. It is all too seldom that a decision can be
arrived at on the basis of the principle used by this businessman,
but, except possibly for the assumption of simple ordering, I know
of no other extralogical principle governing decisions that finds
such ready acceptance.'' \citep[pg. 21]{savage_foundations_1972}
\end{quote}
The idea illustrated is that if $X$ (buying) is preferred to $Y$
(not buying) under condition $A$ (Republican wins) and also under
condition $\neg A$ (Democrat wins), then $X$ is preferred to $Y$.
Savage calls this the sure-thing principle.

Putting it in a more formal way, consider the following three propositions,
$A$, $X$, and $Y$, and let $P(X|A)$, the conditional probability
of $X$ given $A$, represent a measure of a rational belief on whether
$X$ is true given that $A$ is true. Since the measure $P$ requires
rationality of belief, it follows that it satisfies Kolmogorov's axioms
of probability \citep{jaynes_probability_2003,galavotti_philosophical_2005}.
We start with the assumption that 
\[
P(X|A)>P(Y|A).
\]
This expression can be interpreted as stating that if $A$ is true,
then $X$ is preferred over $Y.$ If we also assume
\[
P(X|\neg A)>P(Y|\neg A),
\]
then we obtain, multiplying each inequality by $P\left(A\right)$
and $P\left(\neg A\right)$, respectively, 
\[
P(X|A)P\left(A\right)+P(X|\neg A)P\left(\neg A\right)>P(Y|A)P\left(A\right)+P(Y|\neg A)P\left(\neg A\right).
\]
From $P(A\&\neg A)=1$ and the definition of conditional probabilities
we have
\[
P(X)>P(Y).
\]
In other words, if we prefer $X$ from $Y$ when $A$ is true, and
we also prefer $X$ from $Y$ when $\neg A$ is true, then we should
prefer $X$ over $Y$ regardless of whether $A$ is true. This, of
course, is equivalent to the Savage's example. 

Though STP should hold if agents are making rational decisions, \citet{tversky_disjunction_1992}
and \citet{shafir_thinking_1992} showed that it was violated by decision
makers. In \citet{tversky_disjunction_1992}, they presented several
simple decision-making problems to Stanford students. For example,
in one question students were told about a game of chance, to be played
in two steps. In the first step, not voluntary, players had a 50\%
probability of winning \$200 and 50\% of loosing \$100. In the second
step, a choice needed to be made: whether to gamble once again or
not. If a player accepted a second gamble, the same odds and payoffs
would be at stake. When told that that they won the first bet, 69\%
of subjects said they would gamble again on the second step. When
told they lost, 59\% of the subjects also said they would gamble again.
Following the discussion above, we translate the experiment into the
following propositions. 
\begin{eqnarray*}
A & = & \mbox{"Won first bet,"}\\
\neg A & = & \mbox{"Lost first bet," }\\
X & = & \mbox{"Accept second gamble,"}\\
Y & = & \mbox{"Reject second gamble,"}
\end{eqnarray*}
and we have that 
\[
P(X|A)=0.69>P(Y|A)=0.31,
\]
and
\[
P(X|\neg A)=0.59>P(Y|\neg A)=0.41,
\]
since $P(X)=1-P(Y)$, as $X=\neg Y$. Clearly, $X$ is preferred over
$Y$ regardless of $A$. Later on the semester, the same problem was
presented, but this time students were not told whether the bet was
won or not in the first round. In other words, they had to make a
decision between $X$ and $Y$ without knowing whether $A$ or $\neg A$.
This time, 64\% of students chose to reject the second gamble, and
36\% chose to accept it. Since in this case 
\[
P(X)=0.36<P(Y)=0.64,
\]
there is a clear violation of the STP.

\subsection{Violation of the sure-thing-principle with neural oscillators}

We now turn to the representation of the above situation in terms
of neural oscillators. Following \citet{tversky_disjunction_1992},
we call ``Won/Lost Version'' the case when students were told about
$A$, and the ``Disjunctive Version'' when students were not told
about $A$. Let us start with the Won/Lost Version. In the simplest
case, we have two stimulus oscillators corresponding to the stimuli
associated to the brain representation of $A$ and $\neg A$. Let
us call $s_{a}\left(t\right)$ the oscillator corresponding to $A$,
and $s_{\overline{a}}\left(t\right)$ to $\neg A$. As before, response
oscillators are $r_{1}\left(t\right)$ and $r_{2}\left(t\right)$.
For simplicity, we assume that all oscillators have the same natural
frequency $\omega_{i}=\omega_{0}$. We emphasize that this is not
a necessary assumption, but since the dynamics will lead to synchronization,
this will make the overall computations simpler. Then, when $s_{a}$
is activated, so are the response oscillators, and the dynamics is
given by 
\begin{eqnarray}
\dot{\varphi}_{s_{a}} & = & \omega_{0}-k_{s_{a},r_{1}}^{E}\sin\left(\varphi_{s_{a}}-\varphi_{r_{1}}\right)\nonumber \\
 &  & -k_{s_{a},r_{2}}^{E}\sin\left(\varphi_{s_{a}}-\varphi_{r_{2}}\right)\nonumber \\
 &  & -k_{s_{a},r_{1}}^{I}\cos\left(\varphi_{s_{a}}-\varphi_{r_{1}}\right)\nonumber \\
 &  & -k_{s_{a},r_{2}}^{I}\cos\left(\varphi_{s_{a}}-\varphi_{r_{2}}\right),\label{eq:sa-1}
\end{eqnarray}
\begin{eqnarray}
\dot{\varphi}_{r_{1}} & = & \omega_{0}-k_{r_{1},s_{a}}^{E}\sin\left(\varphi_{r_{1}}-\varphi_{s_{a}}\right)\nonumber \\
 &  & -k_{r_{1},r_{2}}^{E}\sin\left(\varphi_{r_{1}}-\varphi_{r_{2}}\right)\nonumber \\
 &  & -k_{r_{1},s_{a}}^{I}\cos\left(\varphi_{r_{1}}-\varphi_{s_{a}}\right)\nonumber \\
 &  & -k_{r_{1},r_{2}}^{I}\cos\left(\varphi_{r_{1}}-\varphi_{r_{2}}\right),\label{eq:sa-2}
\end{eqnarray}
\begin{eqnarray}
\dot{\varphi}_{r_{2}} & = & \omega_{0}-k_{r_{2},r_{1}}^{E}\sin\left(\varphi_{r_{2}}-\varphi_{r_{1}}\right)\nonumber \\
 &  & -k_{r_{2},s_{a}}^{E}\sin\left(\varphi_{r_{2}}-\varphi_{s_{a}}\right)\nonumber \\
 &  & -k_{r_{2},r_{1}}^{I}\cos\left(\varphi_{r_{2}}-\varphi_{r_{1}}\right)\nonumber \\
 &  & -k_{r_{2},s_{a}}^{I}\cos\left(\varphi_{r_{2}}-\varphi_{s_{a}}\right).\label{eq:sa-3}
\end{eqnarray}
For such a system, it is possible to show \citep[Appendix]{suppes_phase-oscillator_2012}
that a response $x$ is selected when 
\begin{eqnarray}
k_{s_{a},r_{1}}^{E} & = & \alpha x,\label{eq:coupling-asymp-delta-excite-1-A}\\
k_{s_{a},r_{2}}^{E} & = & -\alpha x,\\
k_{r_{1},r_{2}}^{E} & = & -\alpha,\\
k_{r_{1},s_{a}}^{E} & = & \alpha x,\\
k_{r_{2},s_{a}}^{E} & = & -\alpha x,\\
k_{r_{2},r_{1}}^{E} & = & -\alpha,\label{eq:coupling-asymp-delta-excite-6-A}
\end{eqnarray}
and 
\begin{eqnarray}
k_{s_{a},r_{1}}^{I} & = & \alpha\sqrt{1-x^{2}},\label{eq:coupling-asymp-delta-inhibit-1-A}\\
k_{s_{a},r_{2}}^{I} & = & -\alpha\sqrt{1-x^{2}},\\
k_{r_{1},r_{2}}^{I} & = & 0,\\
k_{r_{1},s_{a}}^{I} & = & -\alpha\sqrt{1-x^{2}},\\
k_{r_{2},s_{a}}^{I} & = & \alpha\sqrt{1-x^{2}},\\
k_{r_{2},r_{1}}^{I} & = & 0,\label{eq:coupling-asymp-delta-inhibit-6-A}
\end{eqnarray}
where $\alpha$ is a coupling strength parameter that determines how
fast the solution converges to the phase differences given in (\ref{eq:ideal-phase-diff}).
We have similar equations for $s_{\overline{a}}\left(t\right)$, namely
\begin{eqnarray}
\dot{\varphi}_{s_{\overline{a}}} & = & \omega_{0}-k_{s_{\overline{a}},r_{1}}^{E}\sin\left(\varphi_{s_{\overline{a}}}-\varphi_{r_{1}}\right)\nonumber \\
 &  & -k_{s_{\overline{a}},r_{2}}^{E}\sin\left(\varphi_{s_{\overline{a}}}-\varphi_{r_{2}}\right)\nonumber \\
 &  & -k_{s_{\overline{a}},r_{1}}^{I}\cos\left(\varphi_{s_{\overline{a}}}-\varphi_{r_{1}}\right)\nonumber \\
 &  & -k_{s_{\overline{a}},r_{2}}^{I}\cos\left(\varphi_{s_{\overline{a}}}-\varphi_{r_{2}}\right),\label{eq:sabar-1}
\end{eqnarray}
\begin{eqnarray}
\dot{\varphi}_{r_{1}} & = & \omega_{0}-k_{r_{1},s_{\overline{a}}}^{E}\sin\left(\varphi_{r_{1}}-\varphi_{s_{\overline{a}}}\right)\nonumber \\
 &  & -k_{r_{1},r_{2}}^{E}\sin\left(\varphi_{r_{1}}-\varphi_{r_{2}}\right)\nonumber \\
 &  & -k_{r_{1},s_{\overline{a}}}^{I}\cos\left(\varphi_{r_{1}}-\varphi_{s_{\overline{a}}}\right)\nonumber \\
 &  & -k_{r_{1},r_{2}}^{I}\cos\left(\varphi_{r_{1}}-\varphi_{r_{2}}\right),\label{eq:sabar-2}
\end{eqnarray}
\begin{eqnarray}
\dot{\varphi}_{r_{2}} & = & \omega_{0}-k_{r_{2},r_{1}}^{E}\sin\left(\varphi_{r_{2}}-\varphi_{r_{1}}\right)\nonumber \\
 &  & -k_{r_{2},s_{\overline{a}}}^{E}\sin\left(\varphi_{r_{2}}-\varphi_{s_{\overline{a}}}\right)\nonumber \\
 &  & -k_{r_{2},r_{1}}^{I}\cos\left(\varphi_{r_{2}}-\varphi_{r_{1}}\right)\nonumber \\
 &  & -k_{r_{2},s_{\overline{a}}}^{I}\cos\left(\varphi_{r_{2}}-\varphi_{s_{\overline{a}}}\right),\label{eq:sabar-3}
\end{eqnarray}
and the equivalent equations for (\ref{eq:coupling-asymp-delta-excite-1-A})--(\ref{eq:coupling-asymp-delta-inhibit-6-A}). 

\begin{figure}
\begin{centering}
\includegraphics[width=300pt]{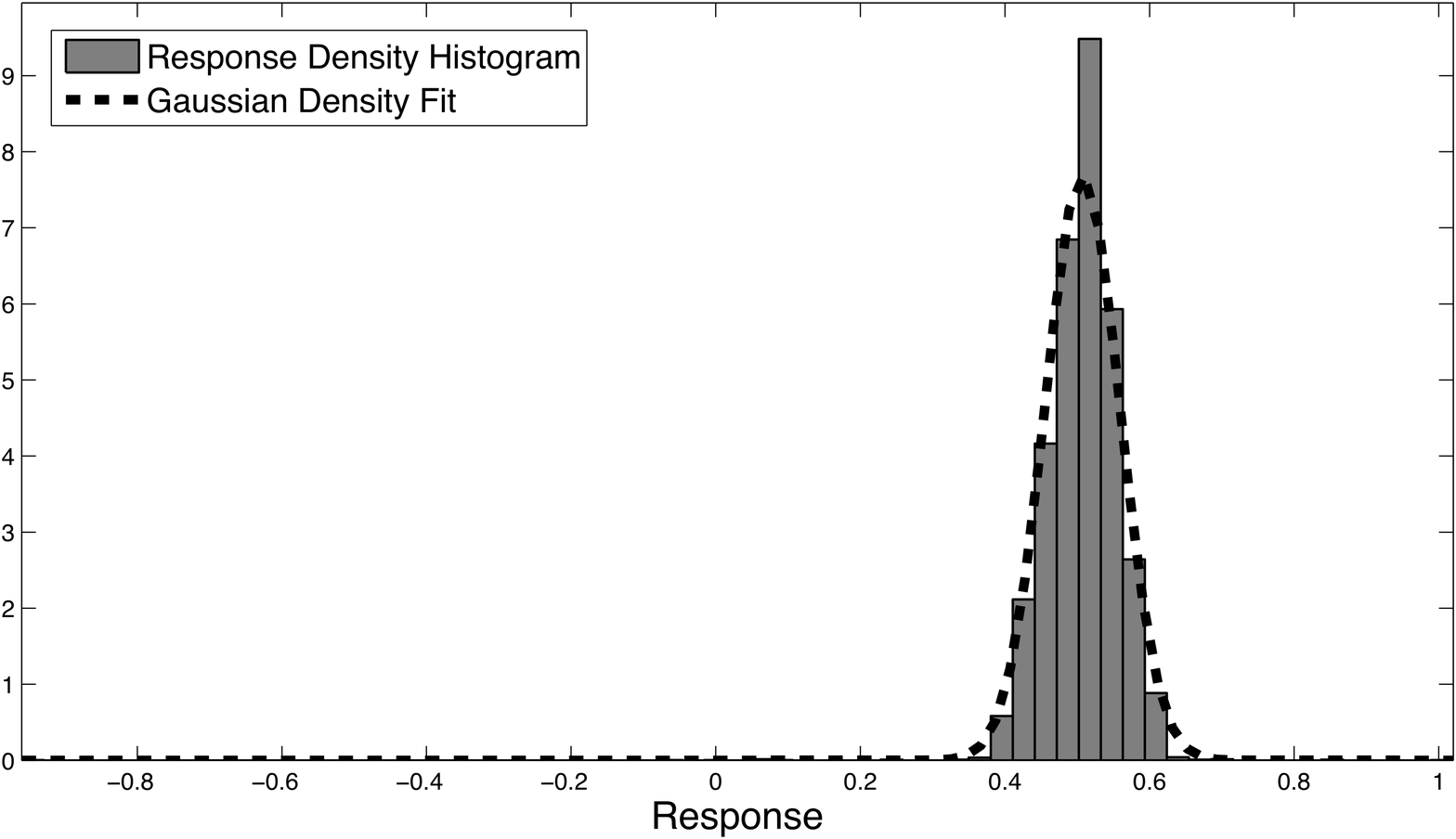} 
\par\end{centering}

\caption{\label{fig:Histogram-density-delta}Histogram density of 300 oscillator
models reinforced using the dynamics given in equations (\ref{eq:learning-phase})--(\ref{eq:learning-asym-inhibitory})
for $y=0.5$. The histogram was obtained using 6300 points corresponding
to 300 oscillators and 21 reinforcement trials per set of oscillators
(trials 40 to 60). All parameters used were the same as those described
in \citet{suppes_phase-oscillator_2012}, and $K_{0}^{'}$ was set
to $4,500$.\textbf{ }The fitted Gaussian has $\mu=0.51$ and $\sigma=0.05$,
with $p<10^{-3}$. }
\end{figure}

Equations (\ref{eq:sa-1})--(\ref{eq:coupling-asymp-delta-inhibit-6-A})
determine a given phase relation for a response $x$ in a continuous
interval, but they do not necessarily model the discrete case of a
selecting $X$ over $Y$. To do so, let us recall that because of
the stochasticity of the initial conditions, if a particular value
$x$ is conditioned through couplings (\ref{eq:coupling-asymp-delta-excite-1-A})--(\ref{eq:coupling-asymp-delta-inhibit-6-A}),
such response is selected according to a smearing distribution $K(x|k_{s_{a}})$.
In fact, if $f\left(y\right)$ is a simple reinforcement given by
a Dirac delta function centered in $z$, i.e. $f(y)=\delta(y-z)$,
then the response distribution $r(x)$ becomes asymptotically \citep{suppes_linear_1959}
\[
r(x)=\int_{-1}^{1}k_{s_{a}}(x|y)\delta(y-z)dy=k_{s}(x|z),
\]
 i.e. $r(x)$ coincides with the smearing distribution at the point
of reinforcement $z$. Thus, by using a reinforcement density given
by a Dirac delta function centered at $z$, we are able to obtain
the shape of the smearing distribution from the oscillator dynamics.
Figure \ref{fig:Histogram-density-delta}
 shows the density histogram for a MATLAB simulation of the three
oscillator model.\textbf{ }Not surprisingly, the model's smearing
distribution $k_{s}(x|z)$ fits well a Gaussian 
\[
k_{s}\left(x|z\right)=\frac{1}{\sigma\sqrt{2\pi}}e^{-\frac{\left(x-z\right)^{2}}{2\sigma^{2}}},
\]
where $\sigma=0.05$. We should mention that the estimation of the
smearing distribution from a reinforcement schedule given by the Dirac
delta function is not realistic for real psychological experiments.
However, because we are running simulations without taking into consideration
other behavioral and environmental aspects, we are able to extract
$k_{s}(x|y)$ from it. 

With the smearing distribution, we may now describe how a stochastic
decision making processes may happen in the brain. Let $X$ and $Y$
be two of the possible responses for a certain behavioral experiment.
In our model, response $X$ would be associated with an oscillator,
$r_{1}\left(t\right)$, and $Y$ with the other, $r_{2}\left(t\right)$.
If $I_{1}=2A^{2}$ and $I_{2}=0$, then the response would clearly
be $I_{1}$. Since $-1\leq x\leq1$, $x$ given by \ref{eq:angle-reinforcement-b},
we may state that $x>0$ corresponds to a response $X$ and $x\leq0$
to response $Y$. If we do this, when the reinforcement is a value
different from $1$ or $-1$, there is a nonzero chance, given by
the smearing distribution, to select each response. In other words,
say a stimulus $s_{a}$ is conditioned to $-1\leq z\leq1$. Then,
\begin{eqnarray*}
P(X) & = & \int_{0}^{1}k_{s_{a}}\left(x|z\right)dx\\
 & \approx & \frac{1}{2}+\frac{1}{2}\mbox{erf}\left(\frac{z}{\sqrt{2}\sigma}\right)
\end{eqnarray*}
and 
\begin{eqnarray*}
P(Y) & = & \int_{-1}^{0}k_{s_{a}}\left(x|z\right)dx\\
 & \approx & \frac{1}{2}-\frac{1}{2}\mbox{erf}\left(\frac{z}{\sqrt{2}\sigma}\right),
\end{eqnarray*}
for $\left|z\right|,\sigma\ll1$, where $\mbox{erf}\left(x\right)$
is the error function. So, we can use the expression 
\[
z=\sqrt{2}\sigma\mbox{erf}^{-1}\left(2P\left(R_{1}\right)-1\right)
\]
to estimate the reinforcement value $z$ leading to the desired probabilities
of responses. Let us summarize the oscillator implementation of the
Won/Lost version. We started with two stimulus, either Won or Lost,
corresponding to the oscillators $s_{a}\left(t\right)$ and $s_{\overline{a}}\left(t\right)$.
Once $s_{a}\left(t\right)$ or $s_{\overline{a}}\left(t\right)$ are
activated, the dynamics of the system leads to a selection of either
$X$ or $Y$, corresponding to the decisions of betting or not betting
on the next step of the game. 

We now model the ``Disjunctive Version.'' Because the response Win
or Lost is not known, we assume that \emph{both} oscillators $s_{a}\left(t\right)$
and $s_{\overline{a}}\left(t\right)$ are activated simultaneously.
Furthermore, since the subject also knows that we either Win or we
Lose, oscillators $s_{a}\left(t\right)$ and $s_{\overline{a}}\left(t\right)$
are incompatible, in the sense that they must represent off-phase
oscillations in an SR-oscillator model. Therefore, the response system
is composed not only of three oscillators, but four: the two stimulus
and the two response oscillators. The equivalent dynamical expressions
for (\ref{eq:sa-1})--(\ref{eq:sabar-3}) when both oscillators are
activated are the following. 
\begin{eqnarray}
\dot{\varphi}_{s_{a}} & = & \omega_{0}-k_{s_{a},r_{1}}^{E}\sin\left(\varphi_{s_{a}}-\varphi_{r_{1}}\right)\nonumber \\
 &  & -k_{s_{a},r_{2}}^{E}\sin\left(\varphi_{s_{a}}-\varphi_{r_{2}}\right)\nonumber \\
 &  & -k_{s_{a},s_{\overline{a}}}^{I}\cos\left(\varphi_{s_{a}}-\varphi_{s_{\overline{a}}}\right)\nonumber \\
 &  & -k_{s_{a},r_{1}}^{I}\cos\left(\varphi_{s_{a}}-\varphi_{r_{1}}\right)\nonumber \\
 &  & -k_{s_{a},r_{2}}^{I}\cos\left(\varphi_{s_{a}}-\varphi_{r_{2}}\right),\label{eq:sdisjunctive-1}
\end{eqnarray}
\begin{eqnarray}
\dot{\varphi}_{s_{\overline{a}}} & = & \omega_{0}-k_{s_{\overline{a}},r_{1}}^{E}\sin\left(\varphi_{s_{\overline{a}}}-\varphi_{r_{1}}\right)\nonumber \\
 &  & -k_{s_{\overline{a}},r_{2}}^{E}\sin\left(\varphi_{s_{\overline{a}}}-\varphi_{r_{2}}\right)\nonumber \\
 &  & -k_{s_{\overline{a}},s_{a}}^{I}\cos\left(\varphi_{s_{\overline{a}}}-\varphi_{s_{a}}\right)\nonumber \\
 &  & -k_{s_{\overline{a}},r_{1}}^{I}\cos\left(\varphi_{s_{\overline{a}}}-\varphi_{r_{1}}\right)\nonumber \\
 &  & -k_{s_{\overline{a}},r_{2}}^{I}\cos\left(\varphi_{s_{\overline{a}}}-\varphi_{r_{2}}\right),\label{eq:sbdisjunctive-1}
\end{eqnarray}
\begin{eqnarray}
\dot{\varphi}_{r_{1}} & = & \omega_{0}-k_{r_{1},s_{a}}^{E}\sin\left(\varphi_{r_{1}}-\varphi_{s_{a}}\right)\nonumber \\
 &  & -k_{r_{1},s_{\overline{a}}}^{E}\sin\left(\varphi_{r_{1}}-\varphi_{s_{\overline{a}}}\right)\nonumber \\
 &  & -k_{r_{1},r_{2}}^{E}\sin\left(\varphi_{r_{1}}-\varphi_{r_{2}}\right)\nonumber \\
 &  & -k_{r_{1},s_{a}}^{I}\cos\left(\varphi_{r_{1}}-\varphi_{s_{a}}\right)\nonumber \\
 &  & -k_{r_{1},s_{\overline{a}}}^{I}\cos\left(\varphi_{r_{1}}-\varphi_{s_{\overline{a}}}\right)\nonumber \\
 &  & -k_{r_{1},r_{2}}^{I}\cos\left(\varphi_{r_{1}}-\varphi_{r_{2}}\right),\label{eq:sdisjunctive-2}
\end{eqnarray}
\begin{eqnarray}
\dot{\varphi}_{r_{2}} & = & \omega_{0}-k_{r_{2},r_{1}}^{E}\sin\left(\varphi_{r_{2}}-\varphi_{r_{1}}\right)\nonumber \\
 &  & -k_{r_{2},s_{a}}^{E}\sin\left(\varphi_{r_{2}}-\varphi_{s_{a}}\right)\nonumber \\
 &  & -k_{r_{2},s_{\overline{a}}}^{E}\sin\left(\varphi_{r_{2}}-\varphi_{s_{\overline{a}}}\right)\nonumber \\
 &  & -k_{r_{2},r_{1}}^{I}\cos\left(\varphi_{r_{2}}-\varphi_{r_{1}}\right)\nonumber \\
 &  & -k_{r_{2},s_{a}}^{I}\cos\left(\varphi_{r_{2}}-\varphi_{s_{a}}\right)\nonumber \\
 &  & -k_{r_{2},s_{\overline{a}}}^{I}\cos\left(\varphi_{r_{2}}-\varphi_{s_{\overline{a}}}\right),\label{eq:sdisjunctive-3}
\end{eqnarray}
with the couplings between $s_{a}$ and $r_{1}$ and $r_{2}$ as well
as $s_{\overline{a}}$ and $r_{1}$ and $r_{2}$ being the same, but
\begin{equation}
k_{s_{a},s_{\overline{a}}}^{I}=\alpha'.\label{eq:inhibit-simultaneous-stimulus}
\end{equation}
As mentioned above, the parameter $\alpha'$ in (\ref{eq:inhibit-simultaneous-stimulus})
represents a measure of the the degree of incompatibility of stimulus
oscillators $s_{a}\left(t\right)$ and $s_{\overline{a}}\left(t\right)$.

Now, as above, let $I'_{1}$ be the mean intensity at $r_{1}\left(t\right)$
given by 
\[
I'_{1}=\left\langle \left(s_{a}(t)+s_{\overline{a}}\left(t\right)+r_{1}(t)\right)^{2}\right\rangle ,
\]
where this time we have both $s_{a}\left(t\right)$ and $s_{\overline{a}}\left(t\right)$
contributing to the intensity. Defining $\delta\varphi_{r_{1},a}\equiv\varphi_{r_{1}}-\varphi_{s_{a}}$
and $\delta\varphi_{r_{1},\overline{a}}\equiv\varphi_{r_{1}}-\varphi_{s_{\overline{a}}}$,
it is straightforward to show that 
\begin{eqnarray*}
I'_{1} & = & A^{2}\left(\frac{3}{2}+\cos\left(\delta\varphi_{r_{1},a}\right)+\cos\left(\delta\varphi_{r_{1},\overline{a}}\right)+\cos\left(\delta\varphi_{r_{1},a}-\delta\varphi_{r_{1},\overline{a}}\right)\right).
\end{eqnarray*}
Similarly, for $r_{2}\left(t\right)$ we have 
\begin{eqnarray*}
I'_{2} & = & A^{2}\left(\frac{3}{2}+\cos\left(\delta\varphi_{r_{2},a}\right)+\cos\left(\delta\varphi_{r_{2},\overline{a}}\right)+\cos\left(\delta\varphi_{r_{2},a}-\delta\varphi_{r_{2},\overline{a}}\right)\right).
\end{eqnarray*}
We notice that, because we now have more oscillators, extra interference
terms shows up in the intensity. We now use the new intensities, $I'_{1}$
and $I'_{2}$ to compute the response, using 
\[
b'=\frac{I'_{1}-I'_{2}}{I'_{1}+I'_{2}}.
\]

At this point, we should remark on the appearance of quantum-like
effects. The model we are using, of interfering oscillators, carry
some similarities with the famous two-slit experiment \citep{suppes_diffraction_1994}.
In the two-slit experiment, a quantum particle can go through two
slits, and then hit a phosphorous screen, thus being detected. Since
a particle is a localized object, going through one slit should mean
no interaction with the other slit, and therefore we should expect
to make no difference whether we keep the other slit closed or open.
However, if we open both slits, we observe different probability patterns
than if we close one of the slits. In other words, if we close one
of the slits, and therefore \emph{know} through which slit the particle
went through, we get a probability of observing the particle in a
certain region of the screen. If, on the other hand, we do not close
one of the slits, we \emph{do not know} through which slit it went,
and when we do not know, the probability of detecting the particle
changes. Formally, let $L$ be the proposition ``went through the
slit on the left because the slit on the right was closed,'' $R=$``went
through the slit on the right because the slit on the left was closed,''
and $A=$``was observed at a certain region on the screen.'' Then,
it is possible to choose a region on the screen such that 
\[
P(A|R)>P\left(\neg A|R\right),
\]
and
\[
P(A|L)>P\left(\neg A|L\right),
\]
i.e., such that there are more particles being detected on it than
outside of it. However, because of quantum interference, if we do
not close the slits, and do not know where they went, it is again
possible to have an $A$ such that the above holds and 
\[
P\left(A\right)<P\left(\neg A\right).
\]
This is the two slit formal equivalent of the violation of STP. 

Now, let us go back to the oscillator model. For the the Win/Lost
Version, a single stimulus oscillator is active at a time. Such oscillator
produces a dynamics that leads to a certain balance between the responses
$X$ and $Y$. This corresponds to one slit open, when there is no
interference from the other slit. However, for the Disjunctive Version,
both oscillators are active, and their activity interferes with each
other, in a way similar to the interference from the two-slit experiment.
The strength of interference, i.e. the degree of coherence between
the two oscillators, is determined by the inhibitory coupling parameter
$k_{s_{a},s_{\overline{a}}}^{I}$. Therefore, when both oscillators
are active, we obtain a probability distribution violating the standard
axioms of probability, and consequently STP. 

To end this section, let us look at a simulation of the violation
of the STP using neural oscillators. We simulated in Matlab R2012a
the oscillators' dynamics determined by equations (\ref{eq:sa-1})--(\ref{eq:sabar-3})
and (\ref{eq:sdisjunctive-1})--(\ref{eq:inhibit-simultaneous-stimulus})
using the Dormand-Prince method. For simplicity, and without loss
of generality, all oscillators were assumed to have the same frequency
of $11$ Hz, as this would allow the use of the couplings given by
(\ref{eq:coupling-asymp-delta-inhibit-1-A})--(\ref{eq:coupling-asymp-delta-inhibit-6-A})
and (\ref{eq:inhibit-simultaneous-stimulus}) without having to invoke
learning equations, as in \citep{suppes_phase-oscillator_2012}. For
our system, the parameters utilized were the following: $\Delta t_{r}=0.2$
s, $\sigma_{\varphi}=\sqrt{\pi/4}$, $k_{s_{a},r_{1}}^{E}=-k_{s_{a},r_{2}}^{E}=k_{r_{1},s_{a}}^{E}=k_{r_{2},s_{a}}^{E}=-0.011$
Hz, $k_{r_{1},r_{2}}^{E}=k_{r_{2},r_{1}}^{E}=-11$ Hz, $k_{s_{\overline{a}},r_{1}}^{I}=-k_{s_{\overline{a}},r_{2}}^{I}=-k_{r_{1},s_{\overline{a}}}^{I}=k_{r_{2},s_{\overline{a}}}^{I}=1$,
$k_{s_{\overline{a}},r_{1}}^{E}=-k_{s_{\overline{a}},r_{2}}^{E}=k_{r_{1},s_{\overline{a}}}^{E}=k_{r_{2},s_{\overline{a}}}^{E}=-0.017$
Hz, $k_{s_{\overline{a}},r_{1}}^{I}=-k_{s_{\overline{a}},r_{2}}^{I}=-k_{r_{1},s_{\overline{a}}}^{I}=k_{r_{2},s_{\overline{a}}}^{I}=1$,
$k_{r_{1},r_{2}}^{I}=k_{r_{2},r_{1}}^{I}=0$, and $k_{s_{a},s_{\overline{a}}}^{I}=k_{s_{\overline{a}},s_{a}}^{I}=-0.011$,
where $\Delta t_{r}$ is the time of response after the onset of stimulus,
and $\sigma_{\varphi}$ is the variance of the initial conditions
for the phase oscillators. We ran 1,000 trials with the above parameter
values, for the Win/Lost Version we obtained $X$ as response 63\%
when the Lost oscillator was used and 58\% with the Win oscillator.
However, for the Disjunctive Version, when the two oscillators were
active, we obtained $X$ as a response 36\% of the runs, clearly showing
a violation of Savage's STP and an interference effect, as expected.

\section{Conclusions}

In this paper we presented a neural oscillator model based on reasonable
assumptions about the behavior of coupled neurons. We then showed
that the predictions of such model presents interference effects between
incompatible stimulus oscillators. Such interference leads to computation
of responses which are inconsistent with standard probability laws,
but are consistent with quantum-like effects in decision making processes
\citet{busemeyer_empirical_2009}. Of course, the fact that we get
interference from $s_{1}$ and $s_{2}$ does not preclude us from
necessarily making rational decisions. We could, for example, modify
the above model, and include an extra oscillator, such that after
reinforcement we could eliminate interference, therefore satisfying
the STP. But our model suggests that a plausible mechanism for quantum
effects in the brain may not need to rely on actual quantum processes,
as advocated by \citet{penrose_emperors_1989}, but instead can be
a consequence of neuronal ``interference.'' 

One interesting feature of the oscillator model is that the amount
of interference between the oscillators is determined by the couplings
between $s_{1}$ and $s_{2}$. The stronger the incompatibility between
the two stimulus, the more we should see interference between the
responses. Furthermore, because interference is a consequence of inhibitory
couplings, we could in principle design experiments where pharmacological
interventions could suppress inhibitory synapses, destroying such
interference effects. 

An attentive reader may question the possibility of modeling decision-making
processes from neurons. In particular, how can we account for what
Pattee claims are non-holonomic constraints not internalized by the
dynamics \citep{pattee_physical_1972,pattee_physics_2001,pattee_epistemic_2012}.
To address this issue, let us examine this claim in more detail. First,
we should notice that there are many examples of non-holonomic constraints
in physical systems that are determined by an underlying and well-understood
dynamics. For example, a ball rolling without slipping on a flat surface
satisfies such type of constraint. But most physicists would not think
that such an example presents fundamental difficulties to the description
of the system ball-surface. In this case, what is a constraint and
what is a physical system is just a matter of computational convenience.
As a trained physicists himself, we believe this is not what Pattee
meant. 

For Pattee, decision making processes require the type of discontinuous
jump that cannot be associated to classical dynamical systems, but
can be thought of as determined by non-holonomic constraints in a
classical dynamics. Since the fundamental laws of physics do not have
such characteristics, he reasons, it remains a mystery how they emerge
from complex systems. He goes on to propose that the jumps come from
the emergence of quantum decoherence (using the modern parlance). 

Going back to the model presented here, the question remains on how
we can derive such emergent characteristics. First, we emphasize that
we are not modeling decision making processes from basic physics,
but instead from emerging properties of collections of neurons. In
other words, we are assuming an emerging dynamics with the necessary
features described by \citet{pattee_physical_1972}: nonlinearity
and time-dependent dynamics \citep{guckenheimer_nonlinear_1983}.
For instance, overall dynamics of a highly complex system of interacting
neurons is modeled mathematically with nonlinear equations describing
the behavior of phases (which adds to the nonlinearity). Furthermore,
the decision-making dynamics changes in time by reinforcements which
are outside of the dynamical system itself \citep{suppes_phase-oscillator_2012}.
Such reinforcements are not only time-dependent, but can also be viewed
as non-holonomic constraints to the dynamics. Thus, in a certain sense,
we are not addressing the fundamental problem posed by Pattee, but
we do push it to another level in the description of neural interactions. 

It is also important to clarify the relationship between our model
and quantum mechanics. We are not claiming that we derive quantum
mechanics from neuronal dynamics. We are showing that certain quantum-like
features \citep[but not all; see ][for details]{de_barros_quantum_2009}
are present in our model. In other words, we start with a classical
stochastic description and obtain some quantum-like results. This
is to be contrasted with Pattee, who argues that classical-like dynamics
emerges from a quantum world, and such emergence happens with the
appearance of life itself%
\footnote{In an earlier paper \citep{suppes_quantum_2007}, we actually discussed
the limits of quantum effects in decision-making processes, and proposed
an experiment where the entanglement of photons could be used to condition
an insect.%
}. 

We end this paper with a comparison between our model and the one
proposed by \citet{khrennikov_quantum-like_2011}. Whereas in Khrennikov's
model the interference of electromagnetic fields account for quantum-like
effects in the brain, in our model we use Kuramoto oscillators to
describe the evolution of coupled sets of dynamical oscillators close
to Andronov-Hopf bifurcations. Because such oscillators obey a wave
equation, they are subject to interference. So, in some sense, we
may say that the oscillator and the field models are equivalent representations
of quantum-like interference in the brain. However, though we can
think of the weak couplings between oscillators originating from the
synapses between neurons, it is also conceivable that the weak oscillating
electric fields generated by sets of oscillators provide sufficient
additional coupling to account for their synchronization and interference.
Thus, we believe that there are not only equivalencies between the
models, but also that our oscillator model might provide an underlying
neurophysiological justification for Khrennikov's quantum-like field
processing. It would be interesting to prove some representation theorem
between neural oscillators and Khrennikov's field model. However,
this seems implausible, since it was shown in \citet{de_barros_joint_2012}
that oscillators may generate processes that are not representable
in terms of observables in a Hilbert space, and Khrennikov's model
relies on a quantum representation of states in terms of Hilbert spaces.

\paragraph*{Acknowledgments}

The author wishes to thank Dr. Gary Oas for critical reading and discussion
on the contents of this manuscript, as well as the anonymous referee
for useful comments and suggestions.

\bibliographystyle{model2-names}
\bibliography{quantumosc}

\end{document}